\documentclass{article}

\usepackage{graphicx}

\begin{document}

\title{Fine tuning of parameters of the universe}

\author{S.G. Rubin \\ Moscow Engineering Physics Institute,  \\
 115409, Moscow, Russia; \\ Centre for CosmoParticle
 Physics "Cosmion"; \\ e-mail: serg.rubin@mtu-net.ru}

\maketitle

\begin{abstract} The mechanism of production of a large number of
universes is considered. It is shown that universes with parameters
suitable for creation of life are necessarily produced as a result
of quantum fluctuations. Fractal structures are formed provided
fluctuations take place near a maximum of the potential. Several
ways of formation of similar fractal structures within our universe
are discussed. Theoretical predictions are compared with
observational data. \end{abstract}


\section{Introduction}
Within many years it was supposed that we live in
a space with Friedman-Robertson-Walker metric. From the astrophysical
point of view it means an expanding universe with small negative
acceleration. From the point of view of modern field theory it
means the vacuum energy density being strictly zero or,
equivalently, a vanishing cosmological constant. There were no clear
theoretical reasons for this, but there was speculation about a hidden
symmetry, implying this strict equality (see, for
example, the review \cite {Star99}).

Four years ago observations \cite {Riess98} indicated some positive
value of the cosmological constant $\Lambda \approx 0.7\rho _M$,
which is only a little less than the average density of matter
$\rho _M$ in the universe. All quantum effects, which give a
contribution to the vacuum energy, surpass this value by many
orders of magnitude. The mechanism of almost complete cancellation
of different contributions is still not understood. And at the same
time, if the cosmological constant would be approximately 200 times
larger as its present value, galaxies would not have been formed
\cite {Wein87} and life would have been impossible. The impression
is that the universe is specially arranged to create life.

The bound of the cosmological constant described above is not the
only case where the existence of life implies a constraint on
parameters in nature. In elementary particle physics there a

re a number of similar examples. I recall here only one - the
smallness of the electron mass. The electron mass is about 2000
times smaller than the nucleon mass. One might suppose that it
would not to matter if it would be several times larger than its
value $0.511 MeV/c^2$. But in this case neutrons would be stable
and the process $p^{+}+e^{-}\rightarrow n+\bar \nu$ would result in
a sharp decrease of proton abundance in the universe with adverse
consequences for the existence of life. We see that the universe is
''adjusted to life'' by a set of parameters and the cosmological
constant is only one of those parameters (for a recent review see
e.g. \cite{Guth00}). It looks like that nature has in store a large
number of universes and only a small number of it is suitable for
life. The question is how to find a mechanism to select those.

In this paper a mechanism of production of a large number of
universes is considered. The universes differ from each other in
physical parameters. It is shown that universes with parameters
suitable for creation of the life are necessarily produced as a
result of quantum fluctuations. The distribution of the universes
has fractal character provided fluctuations take place near
the maximum of a potential. The ways of observation of fractal
structures inside our universe are discussed in Section 3.

\section {Basic postulates} Usually, when a theoretical model is
set up, first a concrete Lagrangian is postulated. Coupling
constants are assumed to be small such that quantum corrections to
the original Lagrangian are considered to be small as well.
Nevertheless, corrections are small only for weak fields, while for
strong fields this does not hold. To be more specific, let us
consider the Lagrangian of a scalar field
\begin {equation}
\label {1}
L = \frac {1} {2} \left ({\partial _ \mu \varphi} \right) ^ 2 -
\frac {{m ^ 2}} {2} \varphi ^ 2 - \frac {\lambda} {4} \varphi ^ 4 ~ .
\end {equation}
One can compute one-loop quantum corrections to the potential and
finds \cite {Linde90}
\begin {equation}
\label {2}
\delta V = \frac {{\left ({3\lambda \varphi ^ 2 + m ^ 2} \right) ^
2 }} {{64\pi ^ 2}} \ln \frac {{\left ({3\lambda \varphi ^ 2 + m ^
2} \right)}} {{2m ^ 2}} - a\varphi ^ 2 - b\varphi ^ 4.
\end {equation}
The last two terms renormalize the mass and coupling constant of
the Lagrangian and depend on the scheme of renormalization. The
first term changes the form of the potential. This is the most
important term for the following considerations. Multi-loop
corrections as well as interaction with other fields may add new
terms to the potential. It is important to note that any simple
interaction causes an infinite number of additional terms to the
original Lagrangian.

It is easy to see, comparing expressions (\ref {1}) and (\ref {2}),
that new terms are small in comparison with the original terms if
$\varphi << m\cdot exp (1/\lambda)$. To get an estimate, one may
choose $m = 100 GeV$, $\lambda = 0.1$, then quantum corrections to
the potential become large at $\varphi \sim 10^{6} GeV$. It is a
rather large energy for an accelerator. However, at an early
inflationary stage of our universe the average value was rather
large, $\varphi > 10^{19} ~ GeV$. Hence, it is necessary to take
into account an infinite number of additional terms in the
Lagrangian (\ref {1}). Moreover, the amplitude of a scalar field is
restricted even more stringently. The logarithm in expression (\ref
{2}) is the result of the summation of an infinite number of terms
\cite {Coleman75}, which converges only when $\varphi < m/\sqrt
{3\lambda}$. Besides, one can see directly from Lagrangian (\ref
{1}) that the interaction term is of order of the mass term when
$\varphi\sim m\sqrt {2 /\lambda}$. Two last estimations are in good
agreement with each other and give a  much smaller value of the
field when quantum corrections are really small. A similar problem
was discussed in the framework of hybrid inflation \cite{Luth99}.

Thus, when considering phenomena in strong fields, i.e.
$\varphi > m/\sqrt {\lambda}$, it is necessary to take into
account all additional terms, inevitably arising due to quantum
corrections. The potential becomes much more complex,
based on the low energy limit of the theory. This can be visualized
by the picture of mountains and valleys. In a mountain area it is
possible to have smooth surfaces with small curvature only in valleys,
i.e., in minima of the potential
energy. After climbing to some height, it becomes obvious
that the shape of the terrain is much more complex.

Usually the potential of interaction of a scalar field is assumed
to be of the most simple form. The property of renormalizability of
the theory is not required if one supposes that gravitational
effects on Plank scale will regularize integrals. Usually, the
fields are weak and quantum corrections are reduced to the
renormalization of parameters of a Lagrangian under the assumption
that the final corrections are small. As consequence of the
previous discussion, at the moment of formation of our universe,
i.e., at large amplitudes of a field, quantum corrections most
likely were comparable with original terms of the Lagrangian, and
its form was much more complex than the Lagrangian considered
above.

The main conclusion is that the choice of any simple form of
Lagrangian with specific parameters leads to difficult problems:
One must explain {\it ab initio} the origin of both the form of
Lagrangian and numerical values of parameters and finally manage to
prove that quantum corrections are small at high energies.
In addition, the field is only a dynamical variable which has no
physical meaning. It is not clear why we should single out the value
$\varphi =0$ when postulating the form of a potential.

Let us take the opposite point of view and limit ourselves
to the minimal number
of specific assumptions about the form of a potential. Namely, let us
postulate some kind of "democracy" - all terms are possible - and
consider consequences of this assumption. More accurately, I
suppose:

\begin{itemize} \item The potential of a scalar field is a
polynomial with an infinite number of terms. Coefficients of
polynomial terms are uncorrelated numbers and are normalized by the
Plank mass $M_{pl}$. As was discussed above, this postulate does
not contradict conclusions of the quantum field theory at low
energy near the bottom of the potential. At high energy it leads to
qualitatively new results.

\item The potential satisfies the conditions $0<V(\varphi )< M _
{pl} ^ 4$($ \hbar = c = 1 $). The first inequality is ordinary one
and is necessary to escape problems with vacuum instability. The
second inequality is needed to avoid influence of effects of
quantum gravity what is usually out of our control and would
completely change all physics. \end{itemize}

As an example, let us consider the following Lagrangian of a scalar
field
\begin {equation}
\label {3}
L = \frac {1} {2} \left ({\partial _ \mu \varphi} \right) ^ 2 -
V (\varphi) ~ .
\end {equation}
The field $\varphi$ is determined in the interval $(-\infty,
+\infty)$. The typical behavior of the potential is represented in
\ref{Fig1}. It should also be stated that the Lagrangian (\ref {2})
is a special case of a more general Lagrangian, in which quantum
corrections to the kinetic term would be taken into account.

The universe is located in one of the minima, where the potential
$V (\varphi)$ can be approximated in a simple way: $V (\varphi)
\approx V (\varphi _ {m}) + a\phi ^ 2 + b\phi ^ 4, \quad \phi =
\varphi - \varphi _ m $. Usually a similar potential is postulated
from the beginning with specific constants $ a $ and $ b $. The
constant $ a $ is connected with mass of a quanta of the field
$\varphi, a = m _ {\varphi} ^ 2 /2$, if $a> 0$. Other universes
occupy other minima which are characterized by a potential with
different parameters $a$ and $b$. The next section is devoted to
cosmological consequences of the above postulates.

\section {Quantum fluctuations as the generator of the universes}
All (quasi) stationary states are located in minima of a potential
and our universe, not being an exception, is located in such a
minimum as well. As there is an enumerable set of minima (remind
that the potential in question is the polynomial with infinite
number of terms), each of which is characterized by some specific
energy density, it seems unlikely that our universe has appeared
just in the minimum with a very small energy density suitable for
life. For an estimate of this probability let us assume that the
probability to end up in a minimum of the potential with energy
density $\rho _ {V} ^ {(m)} = V(\varphi _ m)$ in an interval $d\rho
_ V ^ {(m)}$ is given by
\begin {equation}
dP\left ({\rho _ V ^ {(m)}} \right) = d\rho _V^{(m)} /M_{pl}^4
\end {equation}
(i.e. the uniform distribution of $\rho _ V ^ {(m)}$ is assumed in
the whole interval $(0, M _ {pl}^4)$). The observational value of
energy density in our universe is $\rho _ V \sim 10^{-123}
M_{pl}^4$. Thus, we come to the conclusion that the fraction of
universes with vacuum energy density similar to ours is $\approx
10^{-123}$. It is hard to believe, given an infinite number of such
universes, that an event with such small probability has happened
in nature. We conclude that a mechanism of a "sorting" of the
universes is necessary, and such mechanism  really exists.

\begin{figure} \includegraphics [scale=0.7]{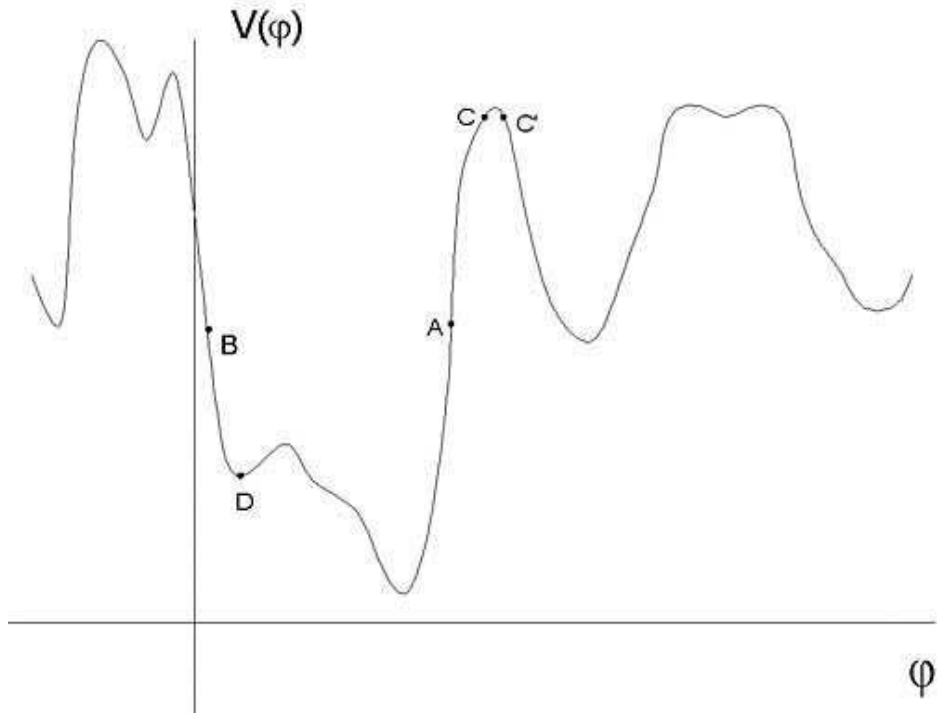}
\caption{\it A part of the potential in a finite range of field
$\varphi$.}\label{Fig1}\end{figure}

To proceed, let us show that if we have a set of potential minima
where life is possible, the field ends up in one of those minima
starting from an arbitrary initial value of the field. Consider
Fig. \ref{Fig1}. Let the field start at a value represented by
point A in Fig. \ref{Fig1}. The spatial area which is
characteristic of the size of fluctuation, is chosen to equal the
Planck scale $(\sim 10^{19}GeV\approx 10^{-33}cm)$. It is a lower
limit where the concept of time can be used. The further destiny of
the area strongly depends on field configurations within this area.
Configurations being important for our considerations are those
where spatial derivatives are small, i.e. $(\partial\varphi _
{\mu})^2 << V(\varphi)$. In this case we can use well developed
methods of inflation theory and especially chaotic inflation \cite
{Linde90}. The inflation paradigm is developing during more than 20
years \cite {Guth81, Kolb91, Khlopov, Brand01}. It successfully
solves the basic problems of cosmology of our universe starting
from the earliest stage since its creation and ending by the stage
of galaxy formation. Here it worth to mention at least the horizon
problem, the flatness problem and the problem of magnetic monopole
absence.


It is important to note that the size of spatial area corresponding
to  fluctuation is growing exponentially. A physical distance $R$
between two points increases like $R(t) \sim exp (Ht)$, while the
size of horizon $1/H$ remains constant. The Hubble parameter $H$ is
connected to the energy density of the potential $H = \sqrt {8\pi
V(\varphi) /3m_{pl}^2}$. It is obvious that an initial causally
connected volume of size $1/H$ is divided into a number $\simeq e^3
$ of causally disconnected areas with the same size $1/H$ in
characteristic time $1/H$. The field values in different areas may
differ from each other due to quantum fluctuations \cite {Rey87}.
It should be noted that this picture represents a look 'from
inside' of the domain. An external observer would detect only field
fluctuations of the size $1/H$ \cite{Guth00,Linde90}. Thus, the
initial area is divided into an increasing number of causally
disconnected areas with various values of the field $\varphi$. This
process is the main process of the inflationary scenario.

Thus, the size of the originally chosen area grows, new areas with
slightly different field values arise inside it. In some of the
areas the field tends to the nearest local minimum of the
potential, while in some other areas field values approach its
local maximum due to fluctuations.

A universe with specific vacuum energy density is formed inside the
domain where the field reached a potential minimum. According to
the above estimate, the probability that this density favors life
of our type, is of the order of $\sim 10^{-123}$. New quantum
fluctuations in any given universe produce new spatial domains with
high energy density $\sim M_{pl}^{4}$ and size $\sim 1/M_{pl}$. The
process of size expansion in these causally disconnected volumes
repeats itself in the manner discussed above \cite{Linde90}. Some
of these domains contain field values corresponding to the slope of
the potential at point B in Fig. \ref{Fig1}. Thus, quantum
fluctuations allow a field in some domains to pass through minima
of the potential.

The destiny of spatial areas where the field overcomes potential
maxima is much more interesting. Consider the fluctuation of the
field near such maximum (point C in \ref{Fig1}). The initial
spatial size of this fluctuation is $\sim 1/H$. When some time of
the order of $\sim 1/H$ has passed this spatial area will be
separated into $e^{3}$ causally disconnected domains with different
field values. The average value of the field $\varphi$ inside some
of these domains will appear at the other side of the maximum
(point $C'$ in Fig. \ref{Fig1}). Each of these domains will be
divided in $e^3$ subdomains of the size of $\approx 1/H$ in time
$1/H$ and some of them will pass back through the maximum of the
potential. This process continuously reproduces itself and already
after several steps a picture of a fractal structure will be
observed. Until now we have not considered the motion of the
classical field which is governed by the classical equation \cite
{Linde90}
\begin{equation} \label{Ecl} \ddot \varphi + 3H\dot \varphi =
-dV/d\varphi ~ . \end{equation}
According to this equation of motion the classical field moves away
from the maximum what could prevent the formation of the fractal
structure. Hence, the development of fractal structure in a final
stage can take place only if the fluctuations are large. More
specifically, let us assume that the classical field changes its
value by $\Delta\varphi_{cl}$ in the time $1/H$. Then a fractal
structure arises if the condition of the fluctuation dominance
$\Delta \varphi_{fluct} >>\Delta \varphi_{cl}$ is satisfied. It
gives enough time for formation of a fractal structure due to the
fluctuations around a maximum. An average value of fluctuations is
well known, given by $\Delta \varphi_{fluct}\simeq H/2\pi$. The
classical motion can be computed explicitly if one approximates the
potential around a maximum by the function
\begin{eqnarray*}
V(\varphi)\simeq V_0 - (\varphi - \varphi_{Max})^2 a^2/2 ~ .
\end{eqnarray*}
An approximate solution of Eq.(\ref{Ecl}) has the form
\begin{eqnarray*} \varphi (t) \simeq \varphi_{Max} + \left[ \varphi
(t=0)-\varphi_{Max}\right] exp\left( \frac{{a^2 M_{pl} }}{{\sqrt
{24\pi V_0 } }}t\right) ~ , \end{eqnarray*}
where the second time derivative is
neglected as is usually done at the inflation stage. The initial field
value $\varphi (t=0)\approx \varphi _{Max}+\Delta\varphi_{fluct}/2$
and the condition of quantum fluctuation dominance is easily found to
be
\begin{equation}
\label{Cond}
\eta \equiv \frac{\Delta \varphi_{fluct}}{\Delta
\varphi_{cl}}\approx H\frac{{2\sqrt {24\pi V_0 } }}{{a^2 M_{pl}
}}>1.
\end{equation}
The number of fractals increases with the parameter the $\eta$. For
an estimate let us take the Planck scale: $V_0 = M_{pl}^{4},
a=M_{pl}$. It leads to the value $\eta \approx 16\pi$ and hence to
a rich fractal structure in the final stage.

It is well known that two domains with field values separated by a
potential maximum, are separated by a wall \cite {Raja82}.
Classically, fields in such domains tend to various (neighboring)
minima, and hence the energy density of the wall grows relative to
the rest of the space. We come to the conclusion that neighbor
universes are separated by field walls with large energy density.

Thus, quantum fluctuations continuously produce spatial domains
with various values of the field $\varphi $. Among these set of
domains it is always possible to find a sequence of domains with
monotonously growing field inside them. The fields in such domains
will consistently pass all minima and maxima on its way.
Eventually, a minimum with energy density suitable for life of our
type will be found.

\section {Fractal structures in our universe} Let us consider in
more detail the process of production of the closed walls. As was
already discussed above, if a causally connected area is placed
near a maximum of the potential, for example $\varphi \geq \varphi_
{max}$, then several subdomains with average field value $\varphi
\leq \varphi_{max}$ will appear inside during the time $1/H$.
Moving along any line, connecting internal and external points of a
subdomain, we necessarily pass through the maximum of the
potential. Therefore, this subdomain is limited by the surface
where the potential has the maximal value, i.e., by the closed
field wall with definite surface energy density. Just after
formation the subdomain is placed near the potential maximum, which
allows to repeat the process. Hence, closed walls of smaller scale
will appear already inside this subdomain. Below it will be shown
that this process results in the formation of fractal structures.

Suppose for a characteristic time $1/H$ several closed walls appear
in a causally connected area of size $R$ near a maximum of the
potential. Denote the number of walls by $N$ and its average size
by $\xi R$, $\xi > 1/e$ ($ \xi \neq 1/e$ due to a possible merging
of causally disconnected subdomains with one common wall). In each
of these subdomains, $N$ new smaller closed walls of size $\xi^{2}
R$ arise during the next time step. Denote by $"a"$ the minimal
size of such a wall that we are able to distinguish. This means
that we may terminate the process after a step $n$ such that $a
\equiv \xi^n R$. The total area of the closed walls in the initial
volume is the sum of areas with closed walls of size greater than
$a$. The simple summation leads to the following result
\begin{equation}
\label {Frac1}
S\approx R ^ 2 q (q ^ n -1) / (q-1),\quad q\equiv \xi ^ 2 N ~ .
\end{equation}
This expression can be written in the form
\begin {equation}
\label {frac2}
S\approx (R/a) ^ D ~ ,
\end {equation}
where $D$ is the fractal dimension. Equating these two expressions,
one obtains
\begin {equation}
\label {Fracdim}
D = 2 + \frac {{\ln \left ({q\frac {{q ^ {\frac {{\ln \left ({a/R}
\right)}} {{\ln \xi}}} - 1}} {{q - 1}}} \right)}} {{\ln (R/a)}} ~ .
\end {equation}
This quantity is constant only when the ratio $R/a$ is large, it is
different for $ q < 1 $ and $ q> 1 $. It can be easily verified
that $D\rightarrow 2$ for $ q < 1$, while for $ q> 1 $,
$D\rightarrow 2 + 3ln (q) /ln(4N)$. To get an estimate, suppose
that the number of closed domains is $ N\approx 4$, and $\xi
\approx 1/e$. The value of the parameter $q$ can be easily
calculated, $q\approx 0.5$. Hence, the fractal dimension of the
system of closed walls $D\approx 2$.

So, if quantum fluctuations lead to the formation of spatial areas
with the field taking a value near a potential maximum, its further
evolution results in a system of enclosing walls. The
characteristic size of the next generations of walls differs from
the previous one approximately by a factor of $e$. The fractal
dimension of such system is $D\approx 2$. The analytical
calculations were done using approximations and hence the
expression (\ref{Fracdim}) has to be considered as an estimate.



According to the above postulates and based on the framework of
chaotic inflation, our universe is a part of a meta-universe which
was formed from one domain surrounded by a closed wall. The
inflationary mechanism provided an exponential increase of its size
from the point of view of an internal observer. The size of the
universe, as presently observed, is estimated to be $\sim 10^{28}
cm$, being smaller by many orders than the characteristic scale of
the meta-universe $\sim 10^{10^{12}}cm$, \cite {Linde91}. Hence,
the walls surrounding our meta-universe are not observable.
Nevertheless, it turns out that the mechanism of generation of
fractal structures appears in a natural way in many models of
inflation. Below three different models which could give rise to
observable consequences are considered.

The first mechanism of the formation of the observable structure is
based on the main postulates of Section 2. Suppose that the
potential $V(\varphi)$ has a local minimum, which is placed close
to the main minimum, as shown in Fig. \ref{Fig1}, point D. The
potential can be approximated as follows
\begin {equation}
\label {Vfrac}
V\left( \varphi \right) = \left\{ \begin {array} {llr} &
\frac{1}{2} m^2 \left( {\varphi - \varphi _ m} \right) ^2;& \left|
{\varphi - \varphi _D} \right| > > \Delta \varphi _D \\
&\frac{1}{2} m^2 \left( {\varphi _ D - \varphi _ m} \right) ^2 +
\frac{1}{2}M^2 \left( {\varphi - \varphi _ D} \right) ^ 2;&\left|
{\varphi - \varphi _D} \right| < < \Delta \varphi _D \\
\end {array} \right. ,
\end {equation}
where $\varphi _ {D}$ is field value at the local minimum.
Inflation takes place when $H >> m$ which is supposed to hold in this
case.

In the vicinity of the local minimum, the equation of motion
(\ref{Ecl}) becomes simpler,
\begin {equation}
\label {Diss}
\ddot \varphi + 3H(\varphi_{D})\dot \varphi + M ^ 2 \left ({\varphi
- \varphi _ D} \right) \approx 0 ~ .
\end {equation}
If $H(\varphi _D)>>M$, dissipation of energy is large and the field
could be located in the local minimum for a long time. We encounter
serious problems, which were discussed in connection with first
inflationary models \cite {Kolb91} where our universe is formed
from the domain in a local potential minimum.

In the case $H(\varphi _D )\leq M$ the situation differs from
the previous one. The field slowly decreases, according to equation
(\ref {Diss}) until it appears in the vicinity of the local minimum
$\varphi = \varphi_{D}$, where the equation of motion can be
reduced to
\begin {equation}
\label {Osc}
\ddot \varphi + M ^ 2 \left ({\varphi - \varphi _ D} \right) \approx 0 ~ ,
\end {equation}
and the total energy of the field is approximately conserved. In
this case the value of the classical field could overcome the local
maximum and approach the nearest deeper minimum of the potential.
In the meantime, the fluctuations described in previous section, occur
in some domains near the local maximum, which leads to the formation
of fractal structure. If this local maximum is deep enough, the
expansion of space increases their sizes not very much. These
fractal structures being small in comparison with the size of our
universe could result in observable consequences.

Let us consider briefly other mechanisms of formation of similar
fractal structures. They are based on a multicomponent or a complex
field instead of a scalar one, which was studied so far. In this
case the mechanism of closed wall production is the same as discussed
above. The only difference is that fluctuations ought to be
investigated near saddle points of the potential rather than near
maxima.

To be specific, let us choose a complex field and, following Ref.\cite
{Freese90} consider the process of formation of our universe in
the framework of natural inflation on the basis of the Lagrangian
\begin {equation}
\label {Vcompl}
L = \partial _ \mu \Phi ^ * \partial ^ \mu \Phi - \lambda \left (
{ \left | \Phi \right | ^ 2 - f/2} \right) ^ 2 - \Lambda ^ 4 \left ({1 -
\cos \theta} \right) ~ ,
\end {equation}
where $\theta$ is a phase of the complex field $\Phi$. The last
term is an approximation of rather complex expression for the
contribution of quantum corrections.

The complex field moves, according to the equations of motion, to a
minimum of the potential at a point $\theta = 0$. At the same time,
due to quantum fluctuations, some part of causally disconnected
domains appears to contain the field value at a saddle point $\theta
= \pi$. In these domains the mechanism of the formation of the
fractal structures sets in. If $\Lambda$ is not very large,
then inflation has no time for a strong increase of the size of the
produced closed walls and we have the opportunity of observation of
fractal structures in our universe.

The last example is based on hybrid inflation, one of the most
promising models of inflation \cite{Linde91a, Dvali94, Luth99}. In
the standard version of hybrid inflation the potential contains two
fields
\begin{equation}
\label{}
V = V_0  + \frac{1}{2}m_\varphi ^2 \varphi ^2  + \frac{1}{2}\lambda
_1 \varphi ^2 \psi ^2  - \frac{1}{2}m_\psi ^2 \psi ^2  +
\frac{1}{2}\lambda _2 \psi ^4 ~ .
\end{equation}
During inflation, the field $\varphi$ rolls down along a valley
$\psi =0$. Just after passing the critical point $\varphi
=m_{\psi}^2/\lambda _1$ the state $\psi =0$ becomes unstable and
field $\psi$ moves (in average) to one of the new stable minima. In
the meantime field fluctuations around the critical point $\psi =0,
\varphi = m_\psi^2 /\lambda_1$ lead to the formation of fractal
structure.



The inflationary mechanisms described above lead to the occurrence
of  fractal structure of the closed walls. After the end of
inflation, as soon as the size of horizon becomes larger than the
characteristic size of closed walls, the walls begin to shrink. The
energy of each wall is proportional to the area of their surface
and concentrates in small spatial domains (in the following they
are considered as pointlike objects)\cite{Ru1}. These high density
clots of energy could serve in the following for star and/or galaxy
formation \cite {Khlopov}. Hence, according to the given models,
the distribution of stars and galaxies should carry fractal
character as well. It is important to note that the total surface
of walls in specific volume is proportional to the total energy
within the volume, while the number of walls is equal to the number
of dense clots.

According to this scenario, it is interesting to find the number of
walls inside a sphere of radius $R$ given by
\begin {equation}
\label {Ntot}
N _ {tot} = \sum _ {i = 1} ^ {n} N ^ {i} = N\frac {N ^ n -1} {N-1}
\approx
\frac {N ^ {n + 1}} {N-1} ~ .
\end {equation}
By analogy with the previous calculations and using
Eq.(\ref{Ntot}), one obtains the distribution  of pointlike dense
objects with fractal dimension $D' \approx lnN/ln (1/\xi)$. For
realistic values $N\approx 4, \quad \xi \approx 1/e$ we find $D'
\approx 1.4$ which differs somewhat from the value $D \approx 2$
previously obtained. This is not surprising because in the first
case we measure the area of surfaces of walls within a certain
volume while in second case we measure the number of walls.

Let us compare our calculations with observational data of spatial
distribution of galaxies and of stars in those galaxies. Recent
data indicate that the distribution of stars and galaxies really
carries fractal character. So, the number of galaxies inside a
sphere of radius $R$ is $N(R)\sim (R)^{2.2\pm 0.2}$ up to the sizes
of 200 Mpc \cite {Labini}.

The distribution of stars inside galaxies also carries fractal
character. In Ref.\cite {FractStar} this fractal dimension was
determined by averaging observational data of ten galaxies and
was found to be equal to $D \sim 2.3$.

Evidently, the observable fractal dimension $D$ in distributions of
stars and galaxies are in agreement with predictions of the given
model. Of course, other mechanisms at a later stage may contribute
to the distribution and change the fractal dimension somewhat, but
the model discussed gives a primordial reason of fractality in the
galaxy and star distribution.

For the sake of completeness it is worth to note another
observational consequence following from the assumption of the
existence of closed walls at an early stage of formation of the
universe. If the mass of a wall is rather large, it can collapse
into a black hole, when shrinking. This process was studied in
Ref.\cite{Ru1}. Hence, the considered model predicts existence of
massive black holes at the centers of galaxies.  This conclusion is
in good agreement with observations. The presence of black holes
with masses of order $10^7 M_{\bigodot}$ at the centers of galaxies
is an established fact by now \cite{Rosen}.

\section {Interaction with fermions}
The interaction of a scalar field with fermions is usually
considered in the form of Yukawa coupling
\begin {equation}
\label {f}
V _ F = g\varphi \bar \psi \psi
\end {equation}
In this case we arrive at a serious problem. The minimum of the
potential, guaranteeing conditions suitable for life, can appear
far from the value $\varphi = 0$. Hence, the term contributing to
the fermion mass $M_F = g \varphi_m $ will be huge comparing with
experimentally observed fermion masses. The problem can be solved
by noticing that the choice (\ref{f}) selects the field value
$\varphi = 0$ which contradicts the main postulates of Section 2.
Let us suppose the interaction has the form
\begin {equation}
\label {f2}
V_F = G (\varphi) \bar \psi \psi,
\end {equation}
which is a generalization of expression (\ref{f}). The function $G
(\varphi)$ is chosen to be a polynomial with random factors in
analogy with the scalar potential $V(\varphi)$. In this case the
fermion mass $M_{F}$ and the constant $g$ of interaction with the
field $\phi = \varphi - \varphi _ {m}$ depend on the number $m$ of
the universe,
\begin {equation} \label {mF} M _ {F} = G (\varphi _ {m}); \quad
\quad g = G ' _ {\varphi } (\varphi _ {m}) . \end {equation}
This expression is obtained by expansion of Eq.(\ref{f2}) in a
power series around the minimum $\varphi_m$. Because we have an
infinite number of universes, it is obvious that for any given
interval of fermion mass $(\mu _ F; \mu _ F + \delta )$ and
function $G (\varphi)$, one can find an appropriate universe such
that the value of the potential at the minimum $V (\varphi _ {m})$
satisfies the equality $\mu _ {F}\cong G(\varphi _ {m})$ with
desired accuracy.

It becomes now possible to use this mechanism for fine tuning of
another parameters of a universe, but not only vacuum energy
density. For example, the existence of life is possible if the
fermion mass lies in an interval $(\mu _{life} , \mu _{life} +
\delta m)$. Then from an infinite set of universes with energy
density suitable for life, one can always choose universes with
suitable values $G(\varphi _{m})$, such that the fermion mass
appears in the given interval. Moreover, this new restricted set of
universes still contains an infinite number of universes and we can
choose a subset of universes with other parameters suitable for
life. Let us introduce a finite set of physical parameters ${\ell
_k}$ which are necessary for creation life in a universe and
enumerable set of universes $\Re (\{ \ell \} _n)$. Here $\{ \ell \}
_n$ is a set of $n$ parameters $\ell _1 ,\ell _2,...,\ell _n$. Then
the process of finding of suitable universe looks like
\begin{eqnarray*}
\Re(\{ \ell \} _0)\Rightarrow \Re(\{ \ell \} _1)\Rightarrow
\Re(\{ \ell \} _2)\Rightarrow ...\Rightarrow \Re(\{ \ell \}
_{N_{life}}) ~ .
\end{eqnarray*}
Here $N_{life}$ is a minimal number of parameters which leads to
conditions suitable for life in the universe. Thus, quantum
fluctuations supply a permanent "search" of universes, suitable for
life by all parameters.

\section {Conclusion} In this paper the mechanism of creation of
universes with given set of microscopic parameters is developed.
The process of formations of each universe is unique, because the
form of potential is unique in the vicinity of each minimum. The
formation of universes is described by different types of
inflationary models. Presently, a large number of models with a
wide range of different potentials are considered as potentially
realistic. Apparently, each of them describes some subset of the
universes of our type. It is shown, that at an early stage of
formation of our universe primordial fractal structures are created
in natural way. Three different scenarios of fractal creation are
considered here. Two of them are based on the well known natural
and hybrid models of inflation. These structures could be the germs
of galaxies and stars. The fractal dimension ($ D\approx 2 $) of
galaxy distribution calculated in the paper is in agreement with
observations.

\section {Acknowledgments} The author is grateful to Professor H.
Kr\"{o}ger for his interest and constant encouragement throughout
the work. This work was partly performed in the framework of
Section "Cosmoparticle physics" of Russian State Scientific
Technological Programme " Astronomy.


\begin{thebibliography}{10}

\bibitem{Star99}
V.Sahny and A.Starobinsky, astro-ph/, astro-ph/9904398  (1999).

\bibitem{Riess98}
A.G.Riess et~al., Astron. Journ. {\bf 116}, 1009 (1998).

\bibitem{Wein87}
S. Weinberg, Phys. Rev. Lett. {\bf 59},  2607  (1987).

\bibitem{Guth00}
A.H.Guth, astro-ph/0002188  (2000).

\bibitem{Linde90}
A.~D. Linde, {\em The Large-scale Structure of the Universe}
(Harwood Academic Publishers, London, 1990).

\bibitem{Coleman75}
S. Coleman,  in {\em Laws of Hadronic Matter} ({\em Ed.:} A.
Zichichi, Academic Press, NY, 1975), p.\ 186.

\bibitem{Luth99}
D.H.Luth, hep-ph/990471  (1999).

\bibitem{Guth81}
A.H.Guth, Phys. Rev.D {\bf 23},  347  (1981).

\bibitem{Rey87}
S.-J. Rey, Nucl. Phys. {\bf B284},  706  (1987).

\bibitem{Raja82}
R. Rajaraman, {\em Solitons and Instantons} (North-Holland
Publishing Company, Amsterdam -- New-York -- Oxford, 1982).

\bibitem{Linde91}
A.~D. Linde, Physica Scripta {\bf T36},  35  (1991).

\bibitem{Kolb91}
E.~W. Kolb, Physica Scripta {\bf T36},  199  (1991).

\bibitem{Freese90}
K. Freese, J.A.Frieman, and A.V.Olinto, Phys.Rev.Lett. {\bf 65},
3233  (1990).

\bibitem{Linde91a}
A.~D. Linde, Phys. Lett. {\bf B259},  38  (1991).

\bibitem{Dvali94} G.Dvali, Q.Shafi, and R.Schaefer, Phys.Rev.Lett.
{\bf 73},  1886 (1994)  .

\bibitem{Khlopov}
M.Yu. Khlopov, {\em Cosmoparticle Physics} (World Scientific,
Singapore-New Jersey-London-Hong Kong, 1999).

\bibitem{Brand01}
R.H.Brandenberger, hep-ph/0101119 (2001)

\bibitem{Labini}
S. Labini, M. Montouri, and L. Pietronero, Phys.Rep. {\bf 293},  61
(1998).

\bibitem{FractStar}
B.G. Elmegreen and D.M. Elmegreen, astro-ph/0012184  (2000).

\bibitem{Ru1}
S.G. Rubin, M.Yu. Khlopov, and A.S. Sakharov, Gravitation \&
Cosmology, Supplement {\bf 6}, 51 (2000).

\bibitem{Rosen}
D.E. Rosenberg and J.R. Rutgers, astro-ph/0012023  (2000).

\end{thebibliography}

\end{document}